\documentclass[aps,eqsecnum]{revtex4}
\usepackage{epsfig}
\usepackage{graphicx}
\usepackage{dcolumn}
\usepackage{bm}
\usepackage{axodraw}
\newcommand{\beq}{\begin{equation}}
\newcommand{\eeq}{\end{equation}}
\newcommand{\bseq}{\begin{subequations}}
\newcommand{\eseq}{\end{subequations}}
\newcommand{\bary}{\begin{eqnarray}}
\newcommand{\eary}{\end{eqnarray}}
\begin{document}
\preprint{ICN/000-03-HEP}
\title{Neutrino absorption by W production in the presence of a magnetic
  field} 
\author{Kaushik Bhattacharya, Sarira Sahu$^\dagger$}
\affiliation{Physics Department, Indian Institute of Technology Kanpur,
Kanppur 208 016, Uttar Pradesh, India\\
\\
$^\dagger$Instituto de Ciencias Nucleares, 
Universidad Nacional Aut\'onoma de M\'exico, Circuito Exterior, C. U., 
A. Postal 70-543, 04510 M\'exico DF, M\'exico\\
and\\ 
Institute of Astrophysics and Particle Physics\\
Department of Physics, 
National Taiwan University, Taipei 106, Taiwan
}
\begin{abstract}
In this work we calculate the decay rate of the electron type
neutrinos into $W$ bosons and electrons in presence of an external
uniform magnetic field. The decay rate is calculated from the imaginary
part of the $W$ exchange neutrino self-energy diagram but in the weak field limit and 
compare our result with the existing one.
\end{abstract}
\maketitle
\section{Introduction}
The topic of neutrino propagation in various non-trivial medium has been
widely studied. Particularly the topic dealing with the passage of neutrinos
through the cosmos or through some compact object containing a magnetic field
is of particular interest. As magnetic fields are ubiquitous in our universe
so when neutrinos travel from their point of creation to their destiny they
may feel the effect of the magnetic fields. As the neutrinos we consider are
standard model chiral neutrinos they do not posses any magnetic moment and the
magnetic field affects them through virtual charged particles inside the loops
of the neutrino self-energy diagrams \cite{Bhattacharya:2004nj,Bhattacharya:2002aj,D'Olivo:2002sp}. It is known that in these circumstances the magnitude
of the magnetic field which is required to affect the neutrino properties is
relatively big, $O(10^{13}){\rm G}$. Such magnitude of magnetic fields are
assumed to exist in neutron stars and magnetars.  Consequently when neutrinos
are produced in the neutron stars their properties can get modified because of
the magnetic field. There has been many attempts to calculate the dispersion
relation of neutrinos in presence of a magnetized medium, which employs the
real part of the neutrino self-energy.  But if we are trying to find out the
absorption rate of the neutrinos then the imaginary part of the neutrino
self-energy becomes important. There has been one attempt in the past
\cite{Erdas:2002wk} where the authors tried to calculate the absorption
coefficient of the neutrinos in a magnetic field background. In their work,
Erdas and Lissia \cite{Erdas:2002wk} calculated the imaginary part of the
self-energy of the neutrino, using the $W$ exchange diagram, and calculated
the absorption coefficient of the neutrino when it decays into an electron
$e^-$ and $W^-$ pair, in presence of a magnetic field.  The present paper is
aimed at calculating the damping rate of the neutrino due to its decay into
$e^-$ and $W^+$ pair, in presence of a magnetic field. The quantities named
absorption coefficient by Erdas and Lissia and damping rate by the present
authors turn out to be the same, both related with the attenuation of a
traveling neutrino in presence of a magnetic field. Our work estimates the
damping rate of neutrinos when the neutrino propagation is damped due to its
decay $\nu_e \to e^- + W^+$ in presence of a magnetic field. The threshold of
the pair production process $\nu_e \to \nu_e + e^- + e^+$ in presence of a
magnetic field has also been calculated \cite{Kuznetsov:1996vy, Dicus:2007gb},
and it is noted that the pair production threshold is smaller than the
neutrino decay threshold. For higher energies of the neutrinos $O(10^{3-4}{\rm
  GeV)}$ and for magnetic field magnitude $O(10^{11-12}{\rm G})$ the neutrino
decay rate becomes a prominent phenomena.  The last reaction is kinematically
forbidden in vacuum but can take place in presence of a magnetic field as in
presence of a magnetic field the dispersion relation of the electron and
positron changes. In their work Erdas and Lissia \cite{Erdas:2002wk}
calculated the imaginary part of the neutrino self-energy using the Feynman
gauge for the $W$ boson propagator. In the present article we calculate the
neutrino self-energy using the unitarity gauge. Our result is higher by 7\%
compared to the one by Erdas and Lissia.

The paper is organized as follows: In sec. 2, we have shown the relation
between the damping rate and the imaginary part of the neutrino self-energy.
In Sec. 3 we explain the terminology of the Schwinger proper time method and
give the expression for the neutrino self energy due to W-exchange. The detail
calculation of the imaginary part of the self-energy in the weak field limit
and the damping rate are evaluated in Sec. 4. The threshold behavior as well
as the asymptotic limit of the damping is also discussed in this section. A
brief conclusion is given in Sec. 5. In appendix A we put some of the
calculational details for the benefit of the reader.

\section{Damping rate of chiral fermions}
\label{damprt}
In this section we briefly describe the main formula of the damping
rate used in this paper. In general, the dispersion relation of a
massless fermion with a momentum $k^\mu=(k^0, \bf{k})$ is not given by
$k^0 = |\bf{k}|$ and in particular, $k^0$ is not zero at zero momentum. Writing
\begin{eqnarray}
k^0=k_r^0 - i\frac{\gamma}{2}\,,
\label{dispri}
\end{eqnarray}
where $k_r^0$ and $\gamma$ being real. The quantity
$M=k_r^0(|\bf{k}|=0)$ is interpreted as an effective mass and $\gamma$
as the fermion damping rate. With the above convention the probability
of the chiral fermion not to decay in a time period $t$ is $\sim
e^{-\gamma t}$. The physically interesting regime is when $\gamma \ll
k_r^0$, since otherwise the system would be over damped and the concept
of a propagating mode is not meaningful. If we decompose the
self-energy of the chiral fermion in real and imaginary parts as
\begin{eqnarray}
\Sigma(k)={\rm Re}\Sigma(k) + i{\rm Im}\Sigma(k)\,,
\end{eqnarray}
the damping rate comes out to be \cite{D'Olivo:1993vm}
\begin{eqnarray}
\gamma = - \frac{1}{k^0}\bar{u}_L\,{\rm Im}\Sigma(k)\,u_L\,,
\label{gamm1}
\end{eqnarray}
where the spinors $u_{L,R}$ satisfy the Dirac equation
\begin{eqnarray}
({\rlap k /}- {\rm Re}\Sigma(k))u_{L,R}= 0\,.
\end{eqnarray}
Eqn.~(\ref{gamm1}) can also be written as
\begin{eqnarray}
\gamma = - \frac{1}{2 k^0_r}{\rm Im} {\rm Tr}\,[{\rlap k/}\Sigma (k)]\,,
\label{gamm2}
\end{eqnarray}
where we treat the spinors as vacuum spinors and do neglect their corrections
in the nontrivial media. In the present work we calculate the imaginary part
of the self-energy of the standard model neutrino in presence of a strong
magnetic field and using that result we find the damping rate of neutrino. 
To compare with earlier work we represent $\gamma$ in units of $m^{-1}$ and
not in the traditional $s^{-1}$.

\section{The $W$ exchange diagram}

\begin{figure}[t!] 
\vspace{0.5cm}
{\centering
\resizebox*{0.6\textwidth}{0.4\textheight}
{\includegraphics{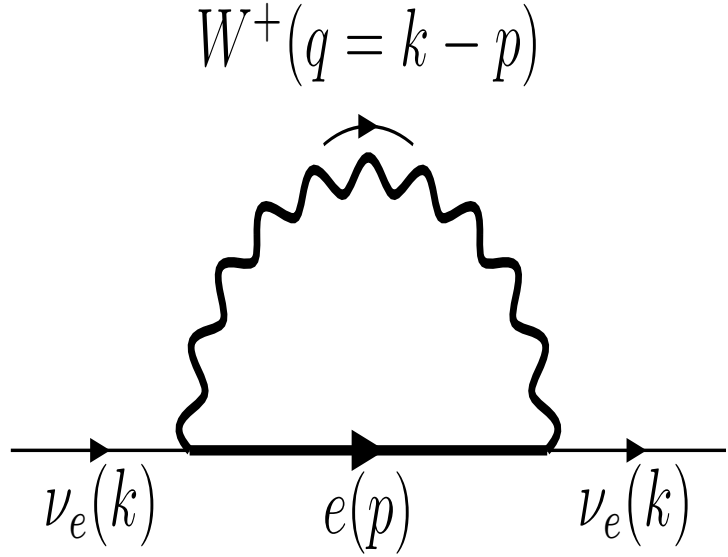}}}
\caption{\small\sf The $W$ exchange diagram. The solid internal virtual lines 
represent modified propagator in presence of an external magnetic field.}
\label{f:ps}
\end{figure}

If the weak SU(2) coupling constant is $g$ and the left-chiral
projection operator is $L\equiv \frac12(1-\gamma_5)$ then the
contribution to the neutrino self-energy from the $W$-boson exchange
diagram is given by,
\beq
-i\Sigma_W (k)=\int \frac{d^4p}{(2\pi)^4}
\left(-i\frac{g}{\sqrt{2}}\right)\gamma_{\mu} L\, i
S_e(p) \left(-i\frac{g}{\sqrt{2}}\right) \gamma_{\nu} L\,\, i W^{\mu\nu}(q) +
{\rm c.t.}
\label{exp1}
\eeq
The c.t. is the counter term which has to be fed in the expression
of the self-energy of the neutrino as the self-energy diagram contains
ultraviolet divergence. The form of the c.t. is such that it cancels
the divergence of the neutrino self-energy in the ${\mathcal B} \to
0$ limit. As the imaginary part of the self-energy in Eq.~(\ref{exp1}) is
relevant for the calculation of $\nu_e \to W^+ + e$ process so the
c.t. will not affect our calculations as it does not contribute to the
imaginary part of the self-energy. Consequently in this article we
will not mention c.t. any further and write the self-energy of the
neutrino in presence of a magnetic field as appearing in
Eq.~(\ref{exp1}) without the counter term.

$S_e(p)$ is the electron propagator in presence of a magnetized
medium and assuming the magnetic field to be along the $z$-axis of the
coordinate system $S_e(p)$is given by:
\beq
iS_e(p)=\int^{\infty}_{0} ds_1\,\, e^{\phi(p,s_1)} G(p,s_1).
\label{slbp}
\eeq
The possible phase term in the electron propagator is not written as it will
disappear in the one-loop calculation of the neutrino self-energy. In
the above equation,
\beq
\phi(p,s_1)=is_1(p_0^2-m^2)-is_1(p_3^2+ p_\perp^2)
\frac{\tan z_1}{z_1}-\epsilon |s_1|,
\eeq
where $m$ is the mass of the electron and we have defined
$z_1=|e|{\mathcal B}s_1$, $p_\perp^2 = p_1^2 + p_2^2$ and $\epsilon$
is an infinitesimal parameter introduced for the convergence of the
integral in Eq.~(\ref{slbp}). Henceforth in this article we will not
require the $\epsilon$ term explicitly and so it will not appear in
further discussions. For the sake of writing the electron propagator in
a covariant form two 4-vectors, $u^\mu$ and $b^\mu$, are used. As the
magnetic field is a frame dependent quantity we assume in the rest
frame of the observer $u^\mu$ is given by,
\begin{eqnarray}
u^\mu = (1, {\bf 0})\,.
\label{u}
\end{eqnarray}
We have a uniform magnetic field along the $z$-axis. Likewise the
effect of the magnetic field enters through the 4-vector $b^\mu$ which
is defined in such a way that the frame in which $u^\mu = (1, {\bf
0})$ we have,
\begin{eqnarray}
b^\mu = (0, \hat{\bf b})\,,
\label{b}
\end{eqnarray}
where we denote the magnetic field vector by ${\mathcal B} \hat{\bf
b}$. With the help of these two 4-vectors the other term appearing
in Eq.~(\ref{slbp}) can be written as,
\begin{eqnarray}
G(p,s_1)=\sec^2 z_1 \left[{\rlap A/} + i {\rlap B/} \gamma_5 +m(\cos^2 z -
i \Sigma^3 \sin z_1 \cos z_1)\right]\,,
\label{gps}
\end{eqnarray}
where $A_\mu$ and $B_\mu$ stand for,
\begin{eqnarray}
A_\mu &=& p_\mu -\sin^2 z_1 (p\cdot u\,\, u_\mu - p\cdot b \,\,b_\mu)\,,
\label{A}\\
B_\mu &=& \sin z_1\cos z_1 (p\cdot u \,\,b_\mu - p\cdot b \,\,u_\mu)\,,
\label{B}
\end{eqnarray}
and
\begin{eqnarray}
\Sigma^3 = \gamma_5 {\rlap /b} {\rlap /u}\,.
\end{eqnarray}
The momentum dependence of the propagator are shown in
Fig.~\ref{f:ps}. The $W$ boson propagator in presence of a magnetic
field is $W^{\mu\nu}(q)$ where $q=k-p$.  In the weak-field limit
i.e. $e{\mathcal B} \ll M_W^2$, the $W$-propagator is given by,
\beq
W_{\mu\nu}(q)=-\frac{1}{(q^2-M_W^2)}\left [ {\rm g}_{\mu\nu}-\frac{1}{M_W^2}
(q_{\mu}q_{\nu}+\frac{ie}{2} F_{\mu\nu})\right ] + \frac{2ie
F_{\mu\nu}}{(q^2-M_W^2)^2}\,.
\eeq
The $q$ dependence of the terms in the denominator on the right hand side of the above
equation can be expressed in integral form as 
\begin{eqnarray}
\frac{1}{(q^2 - M_W^2)} &=& -i \int_0^\infty ds_2 \,\,e^{is_2(q^2 - M_W^2)}\,,
\\
\frac{1}{(q^2 - M_W^2)^2} &=& -\int_0^\infty ds_2 \,s_2\,\,e^{is_2(q^2 - M_W^2)}\,.
\end{eqnarray}
For convergence of the above integrals we require another term similar to
the $\epsilon$ term in Eq.~(\ref{slbp}) which has not been explicitly
written as it will never appear in further discussions. Using
Schwinger's proper time method the contribution to the neutrino
self-energy coming from the $W$ exchange process can be expressed as,
\begin{eqnarray}
\Sigma_W(k) &=& -\frac{ig^2}{2}\int \frac{d^4p}{(2\pi)^4}\,R\,\gamma_{\mu}
S_e (p)\,\gamma_\nu \,L\,W^{\mu \nu}(q)\nonumber\\
&=&-\frac{ig^2}{2}\int \frac{d^4p}{(2\pi)^4}\,
R\,\gamma_{\mu}\int^{\infty}_0 ds_1 \int^{\infty}_0 ds_2\,\,
e^{\phi(p,s_1)}\,G(p,s_1)\, \gamma_{\nu} \,L
\left [
-{\rm g}^{\mu\nu}+\frac{1}{M_W^2} (q^{\mu}q^{\nu}+\frac{ie}{2}
F^{\mu\nu})\right.\nonumber\\
&+& \left. 2es_2F^{\mu\nu} 
\right ]\,e^{is_2(q^2-M_W^2)}\,.
\label{selfen1}
\end{eqnarray}
As these integrals are lengthy and complicated to evaluate, we shall calculate them one by one.
%
\section{ $\Sigma_W(k)$ and the damping rate $\gamma$}
\label{evasigma}
In this section we will evaluate $\Sigma_W(k)$ given in
Eq.~(\ref{selfen1}).  Before writing all the terms in $\Sigma_W(k)$
which will be evaluated separately at first we write down the phase
appearing in the integrand on the right hand side of
Eq.~(\ref{selfen1}). Let us define,
\bary
\Delta &=&\phi(p,s_1)+is_2(q^2-M_W^2)\nonumber\\
&=&-i(s_1 m^2+s_2M_W^2) + i (s_1+s_2) (x_0^2-x_3^2) - i\lambda
(x_1^2+x_2^2) -i\frac{z^2_1}{z_1+z_2}
\frac{{k_{\parallel}}^2}{|e| B}+i\frac{z^2_2}{\lambda}\frac{{k_{\perp}}^2}{|e| B}\,, 
\eary
where,
\begin{eqnarray}
x_0 = p_0 - \frac{k_0 s_2}{s_1+s_2}, \,&&\,
x_3 = p_3 - \frac{k_3 s_2}{s_1+s_2},\\
x_1 = p_1 - \frac{k_1 s_2}{\lambda}, \,\,\,\,\,\,\,\,&&\,
x_2 = p_2  - \frac{k_2 s_2}{\lambda},
\end{eqnarray}
and
\beq
z_i=|e| {\mathcal B}s_i\,,\,\,\,\lambda=s_1\frac{\tan z_1}{z_1}+s_2.
\eeq
Next the expression of $\Sigma_W(k)$ in Eq.~(\ref{selfen1}) is
decomposed into various parts which are evaluated separately. The
decomposition of $\Sigma_W(k)$ is as follows:
\begin{eqnarray}
\Sigma_W(k) = \Sigma^{(1)}_W(k) + \Sigma^{(2)}_W(k) + \Sigma^{(3)}_W(k) 
+ \Sigma^{(4)}_W(k)\,,
\label{sigmabrk}
\end{eqnarray}
where,
\begin{eqnarray}
\Sigma^{(1)}_W(k) &=& \frac{ig^2}{2}\int \frac{d^4p}{(2\pi)^4}\,
\int^{\infty}_0 ds_1 \int^{\infty}_0 ds_2\,\,e^\Delta \,R\,\gamma_{\mu}\,
G(p,s_1)\, \gamma_{\nu} \,L\,{\rm g}^{\mu\nu}\,,
\label{sigm1}\\
\Sigma^{(2)}_W(k) &=& -\frac{ig^2}{2 M_W^2}\int \frac{d^4p}{(2\pi)^4}\,
\int^{\infty}_0 ds_1 \int^{\infty}_0 ds_2\,\,e^\Delta \,R\,\gamma_{\mu}\,
G(p,s_1)\, \gamma_{\nu} \,L\,q^{\mu} q^{\nu}\,,
\label{sigm2}\\
\Sigma^{(3)}_W(k) &=& \frac{e g^2}{4 M_W^2}\int \frac{d^4p}{(2\pi)^4}\,
\int^{\infty}_0 ds_1 \int^{\infty}_0 ds_2\,\,e^\Delta \,R\,\gamma_{\mu}\,
G(p,s_1)\, \gamma_{\nu} \,L\,F^{\mu\nu}\,,
\label{sigm3}\\
\Sigma^{(4)}_W(k) &=& -i e g^2 \int \frac{d^4p}{(2\pi)^4}\,
\int^{\infty}_0 ds_1 \int^{\infty}_0 ds_2 \, s_2\,\,e^\Delta \,R\,
\gamma_{\mu}\,G(p,s_1)\, \gamma_{\nu} \,L\,F^{\mu\nu}\,.
\label{sigm4}
\end{eqnarray}
Doing the Gaussian integrals in the momenta we can write Eq.~(\ref{sigm1}) as,
\begin{eqnarray}
\Sigma^{(1)}_W(k) &=& -\frac{g^2}{(4\pi)^2} \int^{\infty}_0 ds_1
\int^{\infty}_0 ds_2 \, e^{\Delta_0} \frac{1}{(s_1+s_2)\lambda}
R\left [\frac{s_2}{(s_1+s_2)}{\not k}_{\parallel} + i \frac{s_2}{(s_1+s_2)}
(k_0 {\rlap b/} - k_3 {\rlap u/}) \tan z_1 \right. \nonumber \\
&-&\left.\frac{s_2}{\lambda}{\not k}_{\perp}
\sec^2 z_1 \right ] L\,,
\label{sigm1n}
\end{eqnarray}
where,
\begin{eqnarray}
\Delta_0=-i \frac{z^2_2}{(z_1+z_2)} \frac{k^2_{\parallel}}{|e| B} +
i \frac{z^2_2}{\lambda} \frac{k^2_{\perp}}{|e| B}-\frac{i}{|e| B} 
(z_1 m^2+z_2 M^2)\,.
\label{delz}
\end{eqnarray}
The expression in Eq.~(\ref{sigm2}) can also be written as:
\begin{eqnarray}
\Sigma^{(2)}_W(k) &=& -\frac{ig^2}{2 M_W^2}\int \frac{d^4p}{(2\pi)^4}\,
\int^{\infty}_0 ds_1 \int^{\infty}_0 ds_2\,\,e^\Delta \,
\sec^2 z_1 R \left [ 2 q. (A+iB){\not q}-q^2 ({\rlap A/}+i{\rlap B/})
\right ]L\,,\nonumber\\
&=& \Sigma^{(2a)}_W(k) + \Sigma^{(2b)}_W(k)\,,
\end{eqnarray}
where,
\begin{eqnarray}
\Sigma^{(2a)}_W(k) &=&
-\frac{i g^2}{2 M_W^2} \int \frac{d^4p}{(2\pi)^4} \int^{\infty}_0
ds_1 \int^{\infty}_0 ds_2 \,\,e^{\Delta} \sec^2 z_1
R \left[ 2 q.(A+iB){\rlap q/} \right] L\,,
\label{sigm2a}\\
\Sigma^{(2b)}_{W}(k) &=&
\frac{i g^2}{2 M_W^2} \int \frac{d^4p}{(2\pi)^4} \int^{\infty}_0
ds_1 \int^{\infty}_0 ds_2 \,\,e^{\Delta} \sec^2 z_1
R \left[ q^2 ({\rlap A/}+i{\rlap B/})\right] L \,.
\label{sigm2b}
\end{eqnarray}
The elaborate expressions of the above integrals are given in Appendix
\ref{app1}.  The evaluation of $\Sigma^{(3)}_W(k)$ and $\Sigma^{(4)}_W(k)$ are
straight forward and completing the Gaussian integrals over the loop momenta
we obtain:
\beq
\Sigma^{(3)}_W(k)=\frac{g^2}{2M^2} \frac{1}{(4\pi)^2} \int^{\infty}_0
ds_1\int^{\infty}_0 ds_2 
e^{\Delta_0} \frac{s_2}{(s_1+s_2)^2\lambda}
R\left [
i (k_0 {\not b} - k_3 {\not u}) -\tan z_1~{\not k}_{\parallel}
\right ] L\,,
\eeq
and 
\beq
\Sigma^{(4)}_W(k)=\frac{2 eB g^2}{(4\pi)^2} \int^{\infty}_0
ds_1\int^{\infty}_0 ds_2 
e^{\Delta_0} \frac{s_2}{(s_1+s_2)^2\lambda}
R\left [
i (k_0 {\not b} - k_3 {\not u}) -\tan z_1~{\not k}_{\parallel}
\right ] L\,.
\eeq
The damping rate of the neutrino is given by,
\beq
\gamma=-\frac{1}{2 k_0}{\rm Im}~{\cal S}(k)\,,
\label{drate}
\eeq
where
\begin{eqnarray}
{\cal S}(k)={\rm Tr}[{\rlap k/}\Sigma_W(k)]\,.
\label{sk}
\end{eqnarray}
In Eq.~(\ref{drate}) we have used the free dispersion relation of the
neutrinos to write the neutrino energy in the denominator and consequently
here we have $k_r^0 \sim k^0$. The one-loop real
part of the energy will differ from $k_0$ by factors proportional to the Fermi
coupling $G_F$. For simplicity, we can change the integration variables from
$(s_1, s_2)$ to $(z, u)$ where,
\beq
z=\frac{m^2}{eB} (z_1+z_2)= \frac{1}{\beta}  (z_1+z_2)\,, 
\eeq
and
\beq
u=\frac{M_W^2}{m^2}\frac{z_2}{z_1+z_2}= \frac{1}{\eta}\frac{z_2}{z_1+z_2}\,. 
\eeq
With these new variables Eq.~(\ref{delz}) can be written as:
\beq
\Delta_0=-iz\left [1-u\eta+u+u^2\eta\frac{k^2_{\perp}}{M_W^2} 
\left (1-\frac{\beta z}{\Omega}\right)\right]\,,
\label{ndel}
\eeq
where we have defined:
\beq
\Omega=\tan~\beta z(1-u)+\beta z u\eta\,.
\eeq
The phase $\exp(-iz)$ in $\exp(\Delta_0)$ is oscillating and its main
contribution to the $z$ integral will come from $z\le 1$. Also for weak-field
limit the condition is $eB\ll m^2$. So in the weak field limit the product
$\beta z\ll~1$ and we can expand the terms in $\Sigma_W(k)$ up to second order
in $\beta z$ (Appendix \ref{app1} contains the expression of
$\Sigma^{(2)}_W(k)$). This gives:
\beq
\Delta_0\simeq -iz\left [
1-u\eta+u+\frac{1}{3}
\frac{k^2_{\perp}}{m^2} \left( \frac{eB}{M_W^2}\right )^2 u^2 z^2 (1-u\eta)^3 
\right ]\,.
\label{nndel}
\eeq
As the $\Sigma^{i}$, for $i=1,2,3$ and $4$ are in the form:
\beq
\Sigma^{(i)}(k)=f_1 {\rlap k/}_{\parallel(\perp)} + f_2 (k_0 {\rlap b/} - 
k_3 {\rlap u/})\,, 
\eeq
then from Eq.~(\ref{sk}) we see that the trace
\beq
{\rm Tr}[{\rlap k/}(k_0 {\rlap b/} - k_3 {\rlap u/})]=0\,.
\eeq
Consequently the terms which are proportional to $ {\rlap
  k/}_{\parallel(\perp)}$ will only contribute to the damping of the
fermion. Also noting that for a massless neutrino ${\not k}=0$ and
$k^2_{\parallel}=k^2_{\perp}$ we can write ${\cal S}$ in the weak
field limit, where $\beta z\ll~1$, as:
\bary
{\cal S}(k) &\simeq & -\frac{g^2 k^2_{\perp}}{4\pi^2}\int^{\infty}_0 dz
\int^{1/\eta}_0 du \,\,e^{\Delta_0}\nonumber\\
&\times&\left [
\frac{\beta^2 \eta^2 u z (u\eta-1)}{3}
\left \{
2 u^4 z^2 \beta^2 \eta^4-4 u^3 z^2 \beta^2 \eta^3 + u^2(2 z^2\beta^2-1)\eta^2
+ 5 u\eta+2
\right \}\right.\nonumber\\
&-&\left.\frac{i\beta^2\eta^3 u(u\eta-1)}{90}
\left \{
z^2\beta^2 (u\eta-1)^2~(100u^3\eta^3-252 u^2 \eta^2+129 u\eta+8)+
15(8 u^2 \eta^2-22 u\eta+11)
\right \}
\right ]\,.
\label{skexp}
\eary
>From the above equation we can write, 
\beq
{\rm Im}{\cal S}(k)=\frac{(g\beta\eta)^2k^2_{\perp}}{12\pi^2} 
\int^{\infty}_0 dz\int^{1/\eta}_0 du \,u \,(u\eta -1)
\left (
{\cal F}_2 \cos D+{\cal F}_1 \sin D
\right )\,.
\label{imsk}
\eeq
Here
\bary
&&{\cal F}_1 = z(2+5~u\eta-u^2 \eta^2)+2 u^2 \eta^2 \beta^2 z^3 (1-2~u\eta+u^2
\eta^2)\,,\nonumber\\
&&{\cal F}_2 = \frac{1}{2}\eta (11-22~u\eta+8  u^2 \eta^2)+\frac{1}{30}\eta
  (u\eta-1)^2 \beta^2 z^2 (8+129 u\eta-252 u^2\eta^2+100 u^3\eta^3)\,,
\eary
and 
\beq
D=(1+u-u\eta) z+\frac{1}{3} \xi^2 u^2 (1-u\eta)^3 z^3\,,
\eeq
where  we have defined 
\beq
\xi=\frac{k_{\perp}}{m}\left(\frac{e{\cal B}}{M_W^2}\right).
\eeq
By making the substitution 
\beq
z=\frac{y}{(\xi u)^{2/3} (1-u\eta)} 
\eeq
we have
\beq
D=x y + \frac{1}{3} y^3\,,\,\,\,{\rm where}\,\,\,\,
x=\frac{(1+u-u\eta)}{\xi^{2/3} u^{2/3} (1-u\eta)}\,.
\eeq
In the new variable, ${\rm Im}{\cal S}(k)$ can be written as:
\beq
{\rm Im}{\cal S}=-\frac{(g\beta\eta)^2k^2_{\perp}}{12\pi^2\xi^{\frac23}}  
\int^{\infty}_0 dy\int^{1/\eta}_0 du \,u^{\frac13}
\left[{\cal F}_2 \cos \left ( xy+\frac{1}{3} y^3\right )+{\cal F}_1 
\sin \left ( xy+\frac{1}{3} y^3\right )\right]\,.
\label{newsk}
\eeq
In terms of the Airy function:
\beq
Ai(x)=\frac{1}{\pi} \int^{\infty}_0 dy \cos \left ( xy+\frac{1}{3} y^3 
\right )\,,
\label{airy}
\eeq
we can write Eq.~(\ref{newsk}) as:
\beq
{\rm Im}{\cal S}(k)=-\frac{(g\beta\eta)^2k^2_{\perp}}{12\pi\xi^{\frac23}} 
\int^{1/\eta}_0 du\,\,u^{\frac13}
\left [(W_0-W_2~x+W_3) Ai(x) -(W_1-W_3~x) Ai^{\prime}(x)
\right ]\,,
\label{newssk}
\eeq
where
\bary
W_0 &=& \frac{1}{2}\eta (11-22~u\eta+8  u^2 \eta^2)\,,
\label{w0}\\
W_1 &=& \frac{(2+5~u\eta-u^2 \eta^2)}{1+u-u\eta} x\,,
\label{w1}\\
W_2 &=& \frac{\eta
  (u\eta-1)^2 \beta^2  (8+129 u\eta-252 u^2\eta^2+100 u^3\eta^3)}
{30(1+u-u\eta)^2}x^2\,,
\label{w2}\\
W_3 &=& \frac{2 u^2 \eta^2 \beta^2  (1-2~u\eta+u^2\eta^2)}
{(1+u-u\eta)^3}x^3\,,
\label{w3}
\eary
and $Ai^{\prime}(x) \equiv \frac{d Ai(x)}{dx}$.

The numerical integration over $u$ shows that, it saturates much before
reaching the value $1/\eta$, so here $u\eta \ll 1$ and we can neglect the
terms with $u\eta$ or higher power of it in Eq.~(\ref{newssk}). So this gives
\beq
{\rm Im}{\cal S}(k)\simeq
-\frac{(g\beta\eta)^2k^2_{\perp}}{12\pi\xi^{\frac23}} \int^{1/\eta}_0
du\,\,u^{\frac13} 
\left [
\left( \frac{11\eta}{2}-\frac{4}{15} \frac{\eta\beta^2}{(1+u)^2} x^3  \right )
Ai(x) -\frac{2}{1+u} Ai^{\prime}(x)
\right ]\,,
\eeq
with 
\beq
x\simeq \frac{(1+u)}{(\xi u)^{2/3}}.
\eeq
With these the damping rate of the neutrino is given as
\beq
\gamma=\frac{\sqrt{2} G_F k_\perp^2 }{6\pi k_0 \xi^{\frac23}} \left (
  \frac{e{\cal B}}{M_W^2}\right )^2 M_W^2 
\int_0^{1/\eta} du\,\,u^{\frac13} 
\left [
\eta\left( \frac{11}{2}-\frac{4}{15} \frac{\beta^2}{(1+u)^2} x^3  \right )
 Ai(x) -\frac{2}{1+u} Ai^{\prime}(x)
\right ]\,.
\eeq
As $\eta$ is much smaller than one we can henceforth neglect
the part of the integrand proportional to $\eta$ and assume 
$1/\eta \sim \infty$. Also here we assume that $E_{\nu}=k_0\simeq k_{\perp}$ by taking $k_3$ to be very small and  write the above equation as:
\beq
\gamma \sim -
\frac{\sqrt{2}}{3\pi} G_F M^2_W E_{\nu} \left ( \frac{{\cal B}}{{\cal B}_c}\right )^2 
\left ( \frac{m}{M_W}\right )^4 \xi^{-2/3}
\int_0^{\infty} du\,\,\frac{u^{\frac13}}{1+u} Ai^{\prime}(x)\,,
\label{gameqn}
\eeq
where ${\cal B}_c=m^2/e$ is the critical magnetic field.
For the transverse neutrino energy $k_{\perp}$ above the threshold of $e$ and $W$ production, we have observed that the function
\bary
\xi^{-2/3} f(\xi)=
-\xi^{-2/3}
\int_0^{\infty} du\,\,\frac{u^{\frac13}}{1+u} Ai^{\prime}(x) 
&=& \left ( \frac{\xi^{-4/3}}{\sqrt{3} \pi}\right )
 \int^{\infty}_0 \frac{du}{u^{1/3}} K_{2/3} \left (\frac{2}{3} \frac{(1+u)^{3/2}}{\xi u} \right ),
\nonumber\\
&\simeq & 1.  
\label{asyf}
\eary
In Eq.~(\ref{asyf}), we express the
derivative of the Airy function in terms of modified Bessel function to
compare the $u$ dependence of the integrand  with Eq.~(23) of ref.\cite{Erdas:2002wk}, which has
a different $u$ dependence and due to this our result differs from the result by Erdas et al., in the above reference. For neutrino energy much above the threshold, the damping rate behaves as
\beq
\gamma \simeq \frac{\sqrt{2}}{3\pi} G_F M^2_W E_{\nu} \left ( \frac{{\cal B}}{{\cal B}_c}\right )^2 
\left ( \frac{m}{M_W}\right )^4, 
\label{asym}
\eeq
which is same as the absorption coefficient calculated by Erdas et al, in
ref.\cite{Erdas:2002wk}.


\begin{figure}[t!] 
\vspace{0.5cm}
{\centering
\resizebox*{0.4\textwidth}{0.4\textheight}
{\includegraphics{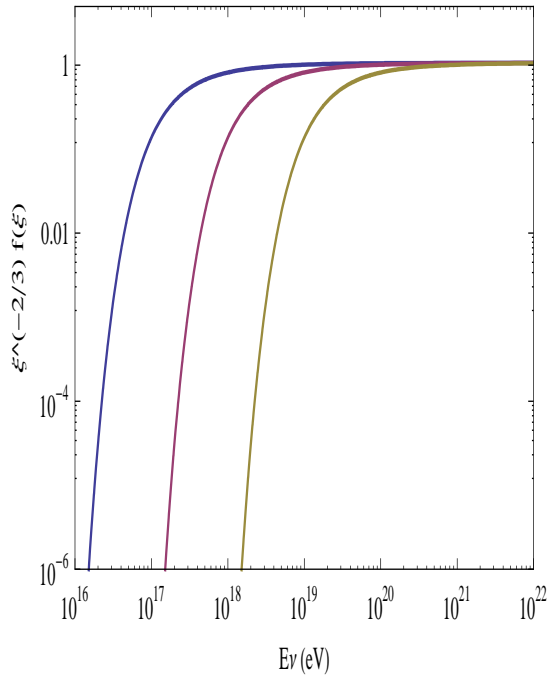}}}
\caption{\small\sf The Eq.~(\ref{asyf}) is plotted as a function of neutrino energy $E_{\nu}=k_{\perp}$ for different ${\cal B}$. From left to right, the curves are for ${\cal B}/{\cal B}_c$ =0.1, 0.01 and 0.001 respectively.}
\label{f2:ps}
\end{figure}

In Fig. 2, we have shown the behavior of Eq.~(\ref{asyf}) as a function of
neutrino energy. For large $k_{\perp}$, it saturates at 1.065 and below the
energy threshold below the function $\xi^{-2/3} f(\xi)$ increases very rapidly
and saturates very fast. Because of this behavior the damping rate also
increases very rapidly from very small value and after crossing the threshold
the behavior becomes linear in neutrino energy.  For small magnetic field, the
saturation energy is larger than the one with larger magnetic field. As can be
seen from the figure, for ${\cal B}/{\cal B}_c$ =0.1, the curve saturates
around $k_{\perp}\simeq 10^{19}\, eV$, whereas for ${\cal B}/{\cal B}_c$
=0.001 the energy is about $10^{21}\, eV$. For the neutrino energy much above
the threshold the asymptotic behavior is given in Eq.~(\ref{asym}), where the
damping rate is proportional to the neutrino energy and also it depends
quadratically on the magnetic field as shown in the above equation. The
behavior can be seen from Fig. 3, where we have plotted the damping rate as a
function of energy for three different values (${\cal B}/{\cal
  B}_c=0.1,\,0.01$ and $0.001$) of the magnetic field. Again larger the
magnetic field, larger is the damping rate and for neutrino energy smaller
than the threshold one, the damping rate is suppressed but increases rapidly
by increasing the energy.

\begin{figure}[t!] 
\vspace{0.5cm}
{\centering
\resizebox*{0.4\textwidth}{0.4\textheight}
{\includegraphics{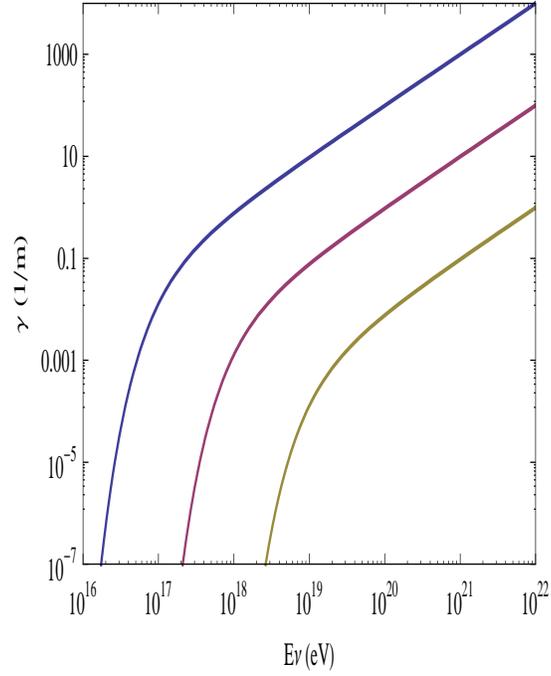}}}
\caption{\small\sf The damping rate is expressed in units of meter inverse is plotted against the transverse energy of the neutrino $E_{\nu}$. From the top to bottom are the curves for $B/B_c=0.1,\, 0.01$ and $0.001$.}
\label{f3:ps}
\end{figure}
\begin{figure}[t!] 
\vspace{0.5cm}
{\centering
\resizebox*{0.4\textwidth}{0.4\textheight}
{\includegraphics{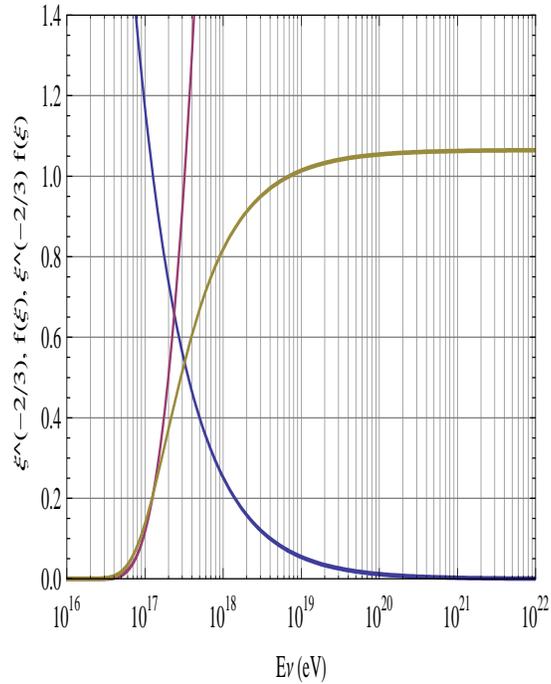}}}
\caption{\small\sf The functions $\xi^{-2/3}$, $f(\xi)$ and $\xi^{-2/3}f(\xi)$ are plotted as
function of $E_{\nu}$ for ${\cal B}/{\cal B}_c=0.1$. The extreme left decreasing curve is the function
$\xi^{-2/3}$, the second increasing curve is $f(\xi)$ and the third one which is saturated around 1 is the function $\xi^{-2/3}f(\xi)$.}
\label{f4:ps}
\end{figure}

The behavior of the process $\nu_e\rightarrow e^- + W^+$ solely depends on $
\xi^{-2/3}f(\xi)$ and the threshold condition satisfies $\xi^{-2/3} f(\xi) <
1$. We have analyzed the threshold behavior of the above process by taking
into account the behavior of the functions $\xi^{-2/3}$, $f(\xi)$ and their
product as function of neutrino energy, which are shown in Fig. 4 for ${\cal
  B}/{\cal B}_c=0.1$ and accurately find out the threshold energy. With
increasing energy the function $\xi^{-2/3}$ decreases and $f(\xi)$ increases
and both of them intersect at a point $\sim 0.66$ which lies before the
saturation point of their product and this is the point of threshold. The
estimated neutrino energy at this point is $E_{\nu}=2.37\times 10^{17}{\rm
  eV}$.  We have observed that, for ${\cal B}/{\cal B}_c=0.01$ and $0.001$,
the intersection point of the two functions as described above remains the
same (0.66) but the neutrino energy shift towards higher values. For ${\cal
  B}/{\cal B}_c=0.01$, we obtain $E_{\nu}=2.37\times 10^{18}{\rm eV}$ and
$2.37\times 10^{19}{\rm eV}$ respectively. For $\nu_{\mu}$ and $\nu_{\tau}$
decay calculations the electron mass has to be replaced by the corresponding
lepton mass.
\beq
\xi = 1.87 \left (\frac{{\cal B}}{{\cal B}_c}\right ),
\eeq
or in terms of the above, we can write the threshold neutrino energy as
\beq
E^{th}_{\nu}=2.36\times 10^{16} \left (\frac{{\cal B}_c}{{\cal B}}\right )\, eV
\eeq
which is about 7\% higher than the one obtained by Erdas et. al.\cite{Erdas:2002wk}.

\section{Conclusion}
\label{conclu}
By using the Schwinger's proper time electron propagator we calculate the
imaginary part of the neutrino self energy in a constant magnetic field
background. In this calculation we assume that the magnetic field to be weak
$e{\cal B}\ll m^2$. From the imaginary part of the neutrino self energy we
calculate the damping rate i.e. conversion of neutrino into electron and
W-boson. We explicitly evaluated all the contributions in the weak field
limit. We found that the behavior of the process $\nu_e\rightarrow e^- + W^+$
solely depends on the quantity $ \xi^{-2/3}f(\xi)$. The threshold energy for
this process can be accurately determined from the intersection of the
functions $\xi^{-2/3}$ and $f(\xi)$ as functions of energy. Also for the
neutrino energy much above the threshold we found that the damping rate is
proportional to the neutrino energy and also it depends quadratically on the
strength of the magnetic field. Our calculation gives an alternative way to
arrive at the absorption cross-section calculated previously in
\cite{Erdas:2002wk}. The damping rate expressed in dimensions of inverse
length matches to a reasonable degree with the old result although the gauge
to fix the weak boson propagator is different in our case. This shows the
gauge invariance of the calculations also.

\appendix\section*{Appendix }
\section{The integral representation of the $W$ exchange diagram}
\label{app1}
The expression in Eq.~(\ref{sigm2}) can also be written as:
\begin{eqnarray}
\Sigma^{(2)}_W(k) &=& -\frac{ig^2}{2 M_W^2}\int \frac{d^4p}{(2\pi)^4}\,
\int^{\infty}_0 ds_1 \int^{\infty}_0 ds_2\,\,e^\Delta \,
\sec^2 z_1 R \left [ 2 q. (A+iB){\not q}-q^2 ({\rlap A/}+i{\rlap B/})
\right ]L\,,\nonumber\\
&=& \Sigma^{(2a)}_W(k) + \Sigma^{(2b)}_W(k)\,,
\end{eqnarray}
where,
\begin{eqnarray}
\Sigma^{(2a)}_W(k) &=&
-\frac{i g^2}{2 M_W^2} \int \frac{d^4p}{(2\pi)^4} \int^{\infty}_0
ds_1 \int^{\infty}_0 ds_2 \,\,e^{\Delta} \sec^2 z_1
R \left[ 2 q.(A+iB){\rlap q/} \right] L\,,
\label{sigm2aa}\\
\Sigma^{(2b)}_{W}(k) &=&
\frac{i g^2}{2 M_W^2} \int \frac{d^4p}{(2\pi)^4} \int^{\infty}_0
ds_1 \int^{\infty}_0 ds_2 \,\,e^{\Delta} \sec^2 z_1
R \left[ q^2 ({\rlap A/}+i{\rlap B/})\right] L \,.
\label{sigm2ba}
\end{eqnarray}
The term $q.(A+iB){\rlap q/}$ in the integrand on the right hand side
of Eq.~(\ref{sigm2aa}) can be written in terms of
$x_0,\,x_1,\,x_2,\,x_3$ as,
\begin{eqnarray}
q. (A+iB){\rlap q/} &=&\left[d_0 - x^2_0 cos^2z_1+\lambda_0 x_0+x^2_1+x^2_2+x^2_3 \cos^2 z_1 +\frac{(x_1k_1+x_2k_2)}{\lambda} \left (s_2-s_1\frac{\tan z_1}{z_1}\right)-\lambda_3 x_3\right]\nonumber\\
&\times & (a_0-\gamma_0 x_0+\gamma_1 x_1+\gamma_2 x_2+\gamma_3 x_3)\,,
\label{sigmp2aa}
\end{eqnarray}
where,
\begin{eqnarray}
a_0=\frac{s_1}{s_1+s_2}{k_{\parallel}}
-\frac{s_1}{\lambda}\frac{\tan z_1}{z_1} k_{\perp}\,,   
\,\,\,
d_0=\frac{s_1 s_2}{(s_1+s_2)^2} {k_{\parallel}}^2 \cos^2 z_1
- \frac{s_1 s_2}{\lambda^2} \frac{\tan z_1}{z_1}  {k_{\perp}}^2\,,
\end{eqnarray}
and,
\begin{eqnarray}
\lambda_0 = \frac{(s_1-s_2)}{(s_1+s_2)}k_0 \cos^2z_1 + i \sin z_1~ \cos z_1
k_3\,,\,\,\,
\lambda_3 = \frac{(s_1-s_2)}{(s_1+s_2)}k_3 \cos^2z_1 + i \sin z_1~ 
\cos z_1 k_0\,.
\end{eqnarray}

where we have assumed that $k_0\simeq k_{\perp}$ by taking $k_3$ to be very small.
Expressing the derivative of the Airy function in terms of Bessel function as 
\beq
Ai^{\prime}(x)=-\frac{1}{\pi} \frac{x}{\sqrt{3}} K_{2/3} 
\left (\frac{2}{3} x^{3/2} \right ),
\label{dampairy}
\eeq
we can express the damping rate as
\bary
\gamma &=&\frac{\sqrt{2}}{3\sqrt{3}} \frac{1}{\pi^2} G_F \frac{k^{2/3}_{\perp}}{k_0} (|e| {\cal B})^{2/3}
(m^2 M_W)^{2/3}
\int^{\infty}_0 \frac{du}{u^{1/3}} K_{2/3} \left (\frac{2}{3} \frac{(1+u)^{3/2}}{\xi u} \right )\nonumber\\
&=& \frac{\sqrt{2}}{3\pi} G_F M^2_W k_{\perp} \left ( \frac{B}{B_c}\right )^2 
\left ( \frac{m}{M_W}\right )^4 
\left ( \frac{1}{\sqrt{3} \pi \xi^{4/3}}\right )
 \int^{\infty}_0 \frac{du}{u^{1/3}} K_{2/3} \left (\frac{2}{3} \frac{(1+u)^{3/2}}{\xi u} \right ),
\label{dampbessel}
\eary
In a similar fashion the term $q^2 ({\rlap A/}+i {\rlap B/})$ in
Eq.~(\ref{sigm2ba}) can be written as,
\begin{eqnarray}
q^2 ({\rlap A/}+i {\rlap B/}) &=&
\left [
       g_0+x^2_0-x^2_1-x^2_2-x^2_3-\frac{2 s_1}{(s_1+s_2)}x_0
       k_0+\frac{2 s_1}{\lambda} \frac{tanz_1}{z_1} (x_1 k_1+x_2 k_2)
+\frac{2 s_1}{s_1+s_2} x_3 k_3
\right]\nonumber\\
&\times& (h+h_0 x_0-\gamma_1 x_1-\gamma_2 x_2-h_3 x_3)\,,
\label{sigmp2ba}
\end{eqnarray}
where,
\begin{eqnarray}
g_0 &=& \left (
\frac{s_2^2}{(s_1+s_2)^2} - \frac{2 s_2}{(s_1+s_2)}\right) {k_{\parallel}}^2
-\left(\frac{s^2_2}{\lambda^2} - \frac{2 s_2}{\lambda}\right ){k_{\perp}}^2\,,
\\
h &=& \frac{s_2}{s_1+s_2} {{\not k}_{\parallel}} \cos^2 z_1 - \frac{s_2}
{\lambda}{{\not k}_{\perp}}+ i\frac{s_2}{s_1+s_2} \sin z_1~\cos z_1 
(k_0{\rlap b/}-k_3 {\rlap u/})\,, 
\end{eqnarray}
and
\bary
h_0 = \cos^2 z_1 {\rlap u/} + i {\rlap b/}  \sin z_1~\cos z_1\,,\,\,\,
h_3 = \cos^2 z_1 {\rlap b/} + i {\rlap u/}  \sin z_1~\cos z_1\,. 
\eary
Using the expressions of $q.(A+iB){\rlap q/}$ and $q^2 ({\rlap A/}+i
{\rlap B/}\,)$ in Eqns.~(\ref{sigm2aa}) and (\ref{sigm2ba}) and doing the
$x_i$ integrals we obtain:
\begin{eqnarray}
\Sigma^{(2a)}_W(k) &=&
-\frac{g^2}{M_W^2}\frac{1}{(4\pi)^2} \int^{\infty}_0
ds_1 \int^{\infty}_0 ds_2 \,\,e^{\Delta_0} \frac{\sec^2 z_1}{(s_1+s_2)\lambda}
\nonumber\\
&\times& R\left [
{\rlap k/}_{\parallel} 
\left \{
\left (\frac{s^2_1 s_2}{(s_1+s_2)^3}k^2_{\parallel} \cos^2 z_1 - 
\frac{s^2_1s_2}{(s_1+s_2) \lambda^2} k^2_{\perp}\frac{\tan z_1}{z_1} 
\right)
-i \left (\frac{2s_1}{(s_1+s_2)\lambda} + \frac{(3s_1-s_2)}{(s_1+s_2)^2}
\cos^2 z_1\right ) 
\right \}\right.\nonumber\\
&&\left.
+{\rlap k/}_{\perp}
\left \{\left (
\frac{s^2_1 s_2}{\lambda^3} \frac{\tan^2 z_1}{z^2_1} {k_{\perp}}^2
-\frac{s^2_1 s_2}{(s_1+s_2)^2\lambda} \frac{\sin z_1~\cos z_1}{z_1}
\right )\right.\right.\nonumber\\
&& \left.\left.+i
\left ( \frac{2 s_1}{(s_1+s_2)\lambda} \frac{\sin z_1~\cos z_1}{z_1} 
+\frac{1}{\lambda^2} (3s_1 \frac{\tan z_1}{z_1}-s_2) 
\right )\right \}
+(k_3 {\rlap u/}-k_0 {\rlap b/}) \frac{\sin z_1~\cos z_1}{s_1+s_2}
\right] L\,,\\
\Sigma^{(2b)}_W(k)&=&
\frac{g^2}{2M_W^2} \frac{1}{(4\pi)^2} 
 \int^{\infty}_0
ds_1 \int^{\infty}_0 ds_2\,\, e^{\Delta_0} \frac{sec^2z_1}{(s_1+s_2)\lambda}
\nonumber\\
&\times& R\left [
{\rlap k/}_{\parallel} 
\left \{
\left (
-\frac{(2s_1+s_2)}{(s_1+s_2)^3}k^2_{\parallel} 
+\frac{s^2_2}{(s_1+s_2)\lambda^2} \left (2 s_1\frac{\tan z_1}{z_1} k^2_{\perp}
\right )\right )\cos^2 z_1
-2 i \cos^2 z_1 \left (
\frac{s_1-s_2}{(s_1+s_2)^2}-\frac{s_2}{(s_1+s_2)\lambda}  \right) 
\right \}\right.\nonumber\\
&&\left. +
{\rlap k/}_{\perp} 
\left \{
\left (-\frac{s^2_2}{\lambda^3} \left (2 s_1\frac{\tan z_1}{z_1} k^2_{\perp}
\right )k^2_{\parallel}+\frac{s^2_2}{\lambda}\frac{(2s_1+s_2)}{(s_1+s_2)^2} 
\right )
-i\left (\frac{2 s_2}{(s_1+s_2)\lambda} + \frac{2s_2}{\lambda^2}
-\frac{2s_1}{\lambda} \frac{\tan z_1}{z_1}
\right )
\right \}\right.\nonumber\\ 
&&\left.
+(k_0{\rlap b/}-k_3 {\rlap u/}) \frac{s_2}{s_1+s_2} \sin z_1~\cos z_1
\left \{
\left ( \frac{2s_1}{s_2(s_1+s_2)}-\frac{2}{(s_1+s_2)} -\frac{2}{\lambda}
\right )\right.\right.\nonumber\\
&&\left.\left.
-i \left (\frac{s_2 (2 s_1+s_2)}{(s_1+s_2)^2}k^2_{\parallel}
-\frac{2 s_1 s_2}{\lambda^2}\frac{\tan z_1}{z_1}
k^2_{\perp}   
\right )\right \}\right ] L\,.
\end{eqnarray}
This results are used to evaluate $\Sigma^{(2)}_W(k)$.

\end{document}